\documentclass[aps,twocolumn,preprintnumbers,%
superscriptaddress,floatfix]{revtex4}
\usepackage[dvips]{graphics}
\usepackage{amsmath,amsfonts,bm}
\usepackage{epsfig}

\newcommand{\be}{\begin{eqnarray}}
\newcommand{\ee}{\end{eqnarray}}
\def\lsim{\mathrel{\rlap{\lower3pt\hbox{\hskip1pt$\sim$}}
     \raise1pt\hbox{$<$}}} 
\def\gsim{\mathrel{\rlap{\lower3pt\hbox{\hskip1pt$\sim$}}
     \raise1pt\hbox{$>$}}} 

\begin{document}
\title{Understanding Dilepton Production in Heavy Ion Collisions\\
by Vector Mesons of Different Varieties}

\author{Gerald E.\ Brown}
\author{Jeremy W.\ Holt}
 \affiliation{ Department of Physics and Astronomy, SUNY, Stony Brook, NY 11794, USA}
\author{Mannque Rho}
 \affiliation{ Institut de Physique Th\'eorique,  CEA Saclay, 91191 Gif-sur-Yvette C\'edex, France}
\date{\today}

\begin{abstract}
A simple schematic model anchored on the notion of hadronic freedom
inferred from hidden local symmetry in the vector manifestation, the
infinite tower of vector mesons in holographic QCD and ``stickiness"
of $\pi\pi$ interactions inferred from dispersion relations is used
to describe the dileptons produced in relativistic heavy ion
collisions at PHENIX/RHIC. It is shown that due to the near
``blindness" of dileptons to Brown-Rho scaling, those dileptons with
invariant mass less than $m_\rho=770$ MeV come mostly from
pion-composites that we interpret as ``transient $\rho$'s."
\end{abstract}

\newcommand\sect[1]{\emph{#1}---}

\maketitle

\sect{I. Introduction:} In a recent article (referred to as
BHHRS)~\cite{bhhrs}, we have proposed a scenario for dilepton
production in heavy ion collisions which is drastically different
from the standard scenario. Starting from the 32 SU(4) mesons which
are massless (in the chiral limit) at $T_c$ in Brown-Rho (BR)
scaling~\cite{br91} and as found in LGS, a period of hadronic
freedom inferred from the theory of hidden local symmetry in the
vector manifestation~\cite{HY:PR} follows. During this period of
free flow the massless hadrons are more or less non-interacting.
They all recover their free-space mass (e.g., $m_\rho=770$ MeV) and
strong interactions at the flash temperature of $T_{flash}\simeq
120$ MeV. We shall refer to these hadrons being ``on-shell" although
they are in a thermal background. At temperature $T\lsim T_{flash}$,
in addition to the $\rho$ mesons coming from the original SU(4)
hadrons of the hadronic freedom region, there will also be those
from $\pi^+ \pi^-$ pairs colliding and forming $\rho$ mesons in the
manner `manufactured' in dispersion theory in the Paris
nucleon-nucleon interaction. The purpose of this Letter is to
describe how the interacting pions contribute to the spectral
function at the dilepton invariant mass $M\lsim m_\rho$.

We shall resort in this paper to an extremely simple schematic model
that borrows the idea from what has been done in treating the
$\rho$-meson exchange in the NN interaction, i.e., the dispersion
relation formulation. A treatment in the framework of hidden local
symmetry generalized to an infinite tower of vector mesons that
arises in holographic QCD~\cite{sakai,mr:kyoto} is proposed as a
field theory approach to the problem.

\begin{figure}[hbt]
\includegraphics[width=4cm]{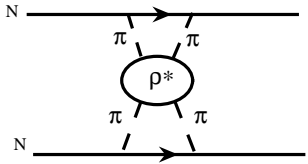}
\caption{The dispersion theory mechanism for binding the $\rho$, and
for obtaining the lower mass pion-composite.} \label{pioncomposit}
\end{figure}
In the Paris potential~\cite{mau,weise}, for instance, the two-pion
exchange potential in the $\rho$ channel (see
Fig.\ref{pioncomposit}) with a helicity amplitude $^3P_0$ can be
characterized by an exchange of a ``pion-composite" with a
distributed mass which shows not only the resonance peak at 770 MeV
but also a non-resonant background. We will refer to this
pion-composite as ``transient $\rho$" or $\rho^*$ for short. Note
that the ``mass" of the $\rho^*$, denoted $m_{\rho^*}$ (equivalent
to $\mu$ in \cite{weise}), can decrease to zero in the chiral limit
but this has {\em nothing} to do with BR scaling. It will be argued
that the largest number of these composites are formed at an energy
\begin{equation}
m_{\rho^*} \simeq m_\rho / \sqrt{2}.
\end{equation}

The way in which these $\rho^*$'s of mass less than $m_\rho$ give
rise to dileptons resembles the way in which $^{12}$C is formed by 3
$\alpha$ particles in stars. In the case of stars, no bound state of
$^8$Be is available to hold two $\alpha$ particles together, but
rather by what we call ``stickiness'' a $0^+$ excited state at 0.092
MeV holds the two $\alpha$ particles together long enough for a
third one to come along and carry the $2\alpha$ system to the 7 MeV
$0^+$ excited state in $^{12}$C. This mechanism is amazingly
effective, the presence of only one excited state in $^8$Be$^*$
being necessary for $10^9$ $\alpha$'s in order to produce the
required amount of $^{12}$C \cite{clayton}.

In the case of the $\rho^*$'s formed by $\pi^+ \pi^-$ collisions, a
``stickiness'' is furnished by the attractive p-wave interaction,
which holds them together long enough so that the ``$\rho^*$'' can
emit dileptons with {\em free-space parameters} (in particular with
$a=2$ and vector dominance). Note the similarity of astrophysical
and heavy ion situations, both of them having a background thermal
field $T$ so that the $^8$Be in the first place and the $\rho^*$ in
the second place form in thermal templates and break up
continuously.

\sect{II. Blindness of Dilepton Production to BR-Scaling
$\rho$-Mesons and   ``Stickiness'' in the $\pi^+ \pi^-$ Collisions:}
We mention that for all practical purposes, dileptons are ``blind''
to BR-scaling $\rho$ mesons, these entering at most at the level of
an order of magnitude less than those from on-shell $\rho$'s and
$\rho^*$'s. This has been established in detail by BHHRS
\cite{bhhrs}.

The ``stickiness'' in $\pi^+ \pi^-$ collisions was used in many
papers on the Paris potential; see e.g.\ \cite{mau}. A lucid review
of the Paris potential can be found in \S 3.10.3 in \cite{weise}.

In dispersion theory the $J=1$ ($^3P_0$) state of $\pi^+$ and
$\pi^-$ experiences a velocity-velocity dependent attraction, which
we take to be linear in the relative velocity. The velocity-velocity
interaction is known to be attractive from the analysis of the $p +
p \rightarrow p +p+\pi^+ +\pi^-$ interaction in which the
interaction between $\pi^+$ and $\pi^-$ can be measured. For the
$\rho$ meson used in the Paris potential the phase shift of the
$\pi^+ \pi^-$ bound state is $\delta = \pi/2$.
(In principle, the phase shift could be greater than $\pi/2$, but in
practice this is close to maximum because the weighting of the
attractive velocity-velocity interaction runs out here.) In the
years of the Mandelstam representation leading towards the Paris
potential the calculations of the helicity amplitudes, the $^3P_0$
one furnishing the properties of the $\rho$, were developed. These
were forgotten, as the quark model took over and the $\rho$ as
discussed was made up out of a $q\bar{q}$ pair. However, the pion
mass would be zero in the chiral limit, and it is clear that the
lowest mass $\rho^*$ is made up chiefly out of $\pi^+$ and $\pi^-$,
or $q^2 \bar{q}^2$ in quark language. The hidden local symmetry is a
field theory with hadronic, not quark, variables so it is clear that
it contains the $\rho^*$ of composite $\pi^+ \pi^-$ nature.

The most numerous $\pi^+ \pi^-$ collisions just following the flash
point where the hadrons go on shell are those with relative angle of
$90^\circ$ because of the large solid angle. These would give a
lower $\delta_{\pi \pi} \sim \pi/8$~\cite{weise}. This is rather far
from a bound state $\rho$ for which $\delta_{\pi \pi} = \pi/2$ is
necessary, but indicates a ``stickiness'' with the $\pi^+$ and
$\pi^-$ pushing each other with their thermal velocities. The pions
only need to be pushed in the right direction so as to stick
together for several fm/c.

In any case,  neither in the case of pions colliding in heavy ion
collisions nor in $^{12}$C production by an additional $\alpha$
attaching itself to a $2\alpha$ state are we dealing with the bound
states of the two particles, but rather with temperature dependent
templates because the two-body systems form and reform as they are
jostled about, due to the background temperature. This feature will
be exploited below for dilepton production.

\sect{III. Making a Quantitative Theory for the PHENIX Dileptons
from the Hidden Local Symmetry Vector Manifestation:} Let us
consider what happens to one batch of SU(4) hadrons, initially at
(nearly) zero mass at $T_c$. From \cite{bhhrs} we have at the flash
point 66 pions, among which there are 18 reconstructed $\rho$'s, 6
of the latter being $\rho^0$ (half of these result from $a_1$
decay). These 6 $\rho^0$ mesons can give rise to dileptons at the
dilepton invariant mass of 770 MeV. From STAR, we
estimate~\cite{BLR-STAR} that there are 6 to 7 $\pi^-$ per $\rho^0$
meson~\cite{footnotepp}, also 6 to 7 $\pi^+$ (corresponding to the
lower and upper limits of number of pions involved in the STAR
ratio); therefore 36 to 49 possible $\pi^+ \pi^-$ collisions. We
will take the mean value $\sim$ 42 in the numerical estimate made
below. These pions have been given off mostly -- among other mesons
-- from the heavy particles, $\rho$ and $a_1$, going on-shell at the
flash point that is reached after a free flow from the hadronic free
regime and are therefore more or less equally distributed over
angle. {\em This spherically symmetrical distribution at the flash
point is the basic structure of the system that we associate with
hidden local symmetry in the vector manifestation} and this will
play a key role in what follows. Thus, there are between 36 and 49
$\pi^+ \pi^-$ collisions per SU(4) batch which can make $\rho^*$'s
with the $\rho^*$'s with the mass 770 MeV identified as the on-shell
$\rho$'s.

The temperature at the end of free flow is between 135 MeV and the
flash temperature, i.e., 120 MeV. Equilibration is assumed to take
place during this interval of temperatures. Ignoring the pion mass
that comes from tiny quark masses, the invariant mass of the
colliding $\pi^+$ and $\pi^-$ is given in terms of the relative
solid angle $\theta$ as
 \be
{\cal M}(\theta, \epsilon)=2\epsilon
\sin\frac{\theta}{2}\label{invariantmass}
 \ee
where $\epsilon$ is the pion energy (or momentum with zero mass).
Since we are neglecting the pion mass, the pion energy is just the
thermal energy $\epsilon \simeq 3T_{flash}$ so that a $\pi ^+$ and
$\pi^-$ colliding head-on will produce a ${\rho}^*$ which is nearly
on-shell, with mass ${\cal M}(\pi, 3T_{flash}) \sim 720$ MeV. We
will call this object $\rho^*_{720}$. By ``head-on collision," we
mean that in which the $\pi^-$ collides within $180 \pm 45^\circ$
with the $\pi^+$, for which the relative solid angle is
\begin{equation}
\frac{\Delta \Omega}{\Omega}=\frac{2\pi \int_{135^\circ}^{180^\circ}{\rm
sine} \theta
  d\theta} {4\pi}\simeq 0.15.
\end{equation}
Since the pions are assumed to get distributed spherically
symmetrically coming from the ``heavy" mesons, the total number of
collisions leading to near on-shell ${\rho}^*$'s can be simply
estimated as $\sim 0.15\times 42 \sim 6$. We note that the number of
these ``near on-shell" $\rho^*$'s is essentially the same as the
on-shell $\rho$'s that would be detected from one set of the $SU(4)$
hadrons at the flash point, 3 of the $\rho^0$'s coming directly from
the set and 3 more from the $a_1$'s decaying into $\rho^0$ plus
$\pi$. This implies that the number of $\rho^0$'s just below the
on-shell $\rho$ peak should be roughly continuous going down in mass
through 770 MeV, the on-shell $\rho$ mass~\cite{footnotep}.

This is, however, {\em not} what is seen in the PHENIX
data~\cite{phenix}. In it the dileptons emitted by the $\rho$ as
well as $\omega$ mesons are shown as a cocktail contribution. This
means that the on-shell $\rho^0$ did not go through the fireball. In
our theory the 3 of the $\rho^0$'s from the $a_1$ decay will have
joined the other $\rho^0$'s in flow and should also be included in
the cocktail peak. This can be done within given errors.

Going down from the cocktail peak
to 720 MeV,  we position our $\sim 720$ MeV $\rho^*$'s at $\sim
1.2\times 10^{-4}$ resulting from 6 $\rho^0$'s. In doing this, we
are equating the number of $\rho^*$'s to that of on-shell $\rho^0$'s
at the point where they join. This defines our normalization.

Now the relative velocity of $\pi^+$ and $\pi^-$ decreases by
$1/\sqrt{2}$ in moving from $\theta=2\pi$ to $\pi$. Therefore from
Eq. (\ref{invariantmass}), we find that at $90^\circ$, the mass of
$\rho^*$ would be $\simeq (720/\sqrt{2})\simeq 509$ MeV. Given our
principal assumption of the equal distribution over angle, there
would be
 \be
f_1 \sim 2.4
 \ee
times more $\pi^+\pi^-$ pairs between 135$^\circ$ and 90$^\circ$
than between 180$^\circ$ and 135$^\circ$ at which $\rho^*$'s are
more or less the on-shell $\rho$'s,  and similarly for 90$^\circ$
and 45$^\circ$ and for 45$^\circ$ and 0$^\circ$. Note that the
larger number of collisions at $\sim 90^\circ$ results from the
larger relative solid angle, and may be characterized as a
continuous function proportional to $\sin\theta$ with relative
velocity $\sin\frac{\theta}{2}$ assumed to give the drop in
$m_{\rho^*}$.

The above reasoning gives the number of $\rho^*$'s as a function of
the solid angle $\theta$ as well as a relation between $m_{\rho^*}$
and the solid angle. All we need for addressing the PHENIX dileptons
is then the dependence of the dileptons on $m_{\rho^*}$.

\sect{IV. Calculating the dileptons from $\rho^*$'s:} In order to
compute the dilepton production from $\pi^+\pi^-$ collisions, we
need to know how the photon couples to the transient $\rho$ (i.e.,
$\rho^*$). As noted earlier, this was parameterized from a knowledge
of the attraction between the two pions determined experimentally
from $p+p\rightarrow p+p+\pi^+ +\pi^-$ reactions. We should note
that we are constructing the $\rho$ and $\rho^*$ as sort of ``energy
fluctuations"; i.e.,  as templates in a heat bath to be weighted
thermally. We are in fact constructing the entire process of
$\pi^+\pi^-\rightarrow \rho^*$ with the maximum
$m^{max}_{\rho^*}=m_\rho=770$ MeV;  the $\rho^*$'s  make the
transition to the observed dileptons.

In a generalized hidden local symmetry theory, the $\rho^*$ coupling
to the virtual photon is vector-dominated by the infinite tower of
vector mesons as is suggested in holographic dual QCD
(hQCD)~\cite{sakai,mr:kyoto}. When
the $\pi^+$ and $\pi^-$ have exactly opposite momenta, the on-shell
$\rho$, here identified at $\sim$ 720 MeV corresponding to the flash
temperature, will be produced. This corresponds to the lowest-lying
vector meson $\rho (770)$ with width of 150 MeV in the tower to
which the dileptons with invariant mass peaked at 770 MeV will
couple. These dileptons are counted as cocktails in PHENIX. The
$\rho^*$'s with mass less than 770 MeV that will be produced at
lower angles will couple to dileptons through the higher-lying
members of the infinite tower. We can assume using closure
approximation that the photon couples {\em point-like} to the latter
since the closure energy $\bar{E}$ should be much greater than the
mass of the $\rho^*$'s lying below 720 MeV. This point-like photon
coupling to $\rho^*$'s via the massive members of the tower will be
the key ingredient for allowing the simple numerical estimates to be
made below.
%

\begin{figure}[hbt]
\includegraphics[width=6.5cm, angle=-90]{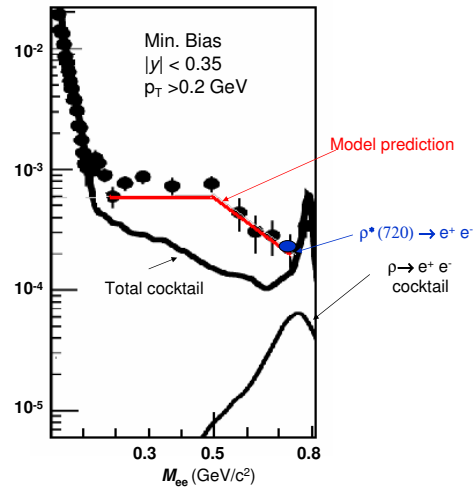}
\caption{PHENIX data $(1/N_{evt})dN/dM_{ee} (c^2/{\rm GeV})$ in
PHENIX acceptance vs. $M_{ee}$.} \label{phenix}
\end{figure}

Now in the $e^+e^-$ CM frame, the photon propagator in the dilepton
coupling to the $\rho^*$ is
 \be
D_\gamma \sim 1/{{\cal M} (\theta)}^2\label{photonprop}
 \ee
where ${\cal M}$ is defined in Eq.(\ref{invariantmass}). $\epsilon$
will be taken to be $3T_{flash}$, so will be omitted in the
expression in what follows. A simple calculation shows that the
propagator (\ref{photonprop}) gives in the diletpon cross section a
scaling factor
 \be
f_2\equiv \left(\frac{{\cal M} (\pi)}{{\cal
M}(\theta)}\right)^2\approx \left(\frac{m_{\rho^*_{720}}}{{\cal
M}(\theta)}\right)^2
 \ee
Thus, given the assumption of the direct coupling explained above,
we would expect, in going from $m_{\rho^*}\simeq 720$ to $\sim 509$
MeV, an increase in $\rho^*$'s of a factor
 \be
 f_1 f_2 \sim 2.4\times 2 \sim 4.8.
 \ee
This increase in number of $\rho^*$'s from 720 MeV to 500 MeV --
which is expected to be more or less linear -- is approximated by a
straight line in Fig.\ref{phenix}.

Now in going from $\sim$ 500 to $\sim$ 200 MeV, one would expect the
decrease in the number of $\pi^+\pi^-$ collisions to be roughly
canceled by the factor $({\cal M} (\pi/2)/{\cal M}(\theta))^2$ so
that the number of experimental points to be more or less flat. This
is indicated as a rough approximation by a straight horizontal line
in Fig.\ref{phenix}. Note that our model does not include the fact
that the $\rho^*$ mass cannot drop below $2m_\pi$.

To compare with experiments, the cocktail contributions need to be
included. Even without the latter, the agreement with experiments is
quite satisfactory. We expect that the same mechanism will be
operative for the NA60 dileptons except that much fewer pions will
be involved here.

We suggest that the rapid rise at very low masses other than that
from $\pi^0$ decay comes from the pions going into $\sigma$-mesons.
(See Fig.5 of Shuryak and Brown~\cite{shuryak-brown}.) The pions
that result from the decay of the $\sigma$ have non-thermal, very
low kinetic energies. This matter will be considered in a separate
paper.
\begin{figure}[tb]
\begin{center}
\includegraphics[width = 3cm, angle=-90]{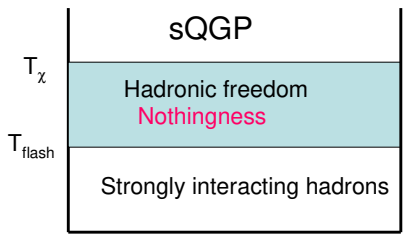}
\caption{The proposed temperature phase diagram modified by the
hidden local symmetry in the vector manifestation that is
conjectured to lead to Hadronic Freedom.  One expects a similar
modification in density.} \label{phasediagram}
\end{center}
\end{figure}

\sect{Conclusions} Hidden local symmetry in the vector manifestation
which gives rise to BR scaling suggests hadronic freedom between the
critical point and the flash point. The way in which the 32 $SU(4)$
hadrons massless at $T_c$ according to the calculations of Park, Lee
and Brown starting from unquenched QCD~\cite{PLB}  go on-shell at
the flash point leads to the ``blindness'' of the theory to
BR-scaling $\rho$ mesons.

We have reconstructed the $\pi^+\pi^-$ pairs which experience
attraction and which are identified with the usual on-shell $\rho$
mesons when their phase shift $\delta_{\pi\pi}$ is equal to $\pi/2$
as in the construction of the Paris potential for NN scattering.
When a bound state is not formed, there is still enough attraction
in the lower-mass fluctuation $\rho^*$ in order to hold the $\pi^+$
and $\pi^-$ together long enough to produce dileptons. These
$\rho^*$'s coupled to the infinite tower of $\rho$ mesons are
sufficient to explain the low-mass dileptons in the PHENIX
experiment.

Our principal conclusion is that the PHENIX data can be successfully
analyzed in terms of hidden local symmetry in the vector
manifestation together with the role of the infinite tower suggested
in hQCD~\cite{sakai,mr:kyoto}. This result follows from the
blindness of the dileptons to BR-scaling vector mesons, the concept
of the flash point and hadronic freedom, all of which are an
essential ingredient of HLS/VM. Although not observable directly, BR
scaling plays a crucial role in understanding the observed spectrum.
The BR-scaling vector mesons that carry signals for partial as well
as full chiral restoration should in principle be present below the
$\rho$ peak but are highly suppressed by the vector manifestation of
hidden local symmetry, rendered invisible in the background of the
pion-composites.

If the proposed scenario is correct as suggested by our analysis,
then the standard phase diagram may have to undergo a major
revamping. There have been suggestions in the literature that the
presently ``predicted" phase structure at high density and low
temperature could be wrong. In fact there can be a plethora of other
phases that could replace or render obsolete the ones figuring in
the present phase diagram: Kaon condensation at a density as low as
3 times normal nuclear matter which would send compact stars into
black holes at higher density~\cite{BLR-kaon}, a half-skyrmion
quantum critical phase related to hadronic freedom in density below
the chiral critical density~\cite{halfskyrmion}, a quarkyonic phase
with quark confinement but restored chiral
symmetry~\cite{quarkyonic}, an anomaly-induced spectral continuity
with no phase transition~\cite{yamamoto} etc. At present, there is
no model-independent theoretical method to determine which are
viable and which are not. On the contrary, it has been generally
accepted, based on available lattice QCD and experimental data, that
the phase structure at high temperature and low-density is more or
less known modulo some details. The present study, however, suggests
that this understanding is not entirely correct. The dilepton
experiments indicate that there is the hitherto unsuspected
``Hadronic Freedom" phase shown in Fig.\ref{phasediagram} which
makes a drastically different picture of how phase transitions occur
in temperature (as suspected in density).

\begin{acknowledgments}
We are grateful for helpful correspondences with M. Harada and C.
Sasaki. This work was partially supported by the U.\ S.\ Department
of Energy under Grant No.\ DE-FG02-88ER40388 and the U.\ S.\
National Science Foundation under Grant No.\ PHY-0099444.
\end{acknowledgments}




\begin{thebibliography}{99}


\bibitem{bhhrs} G. E. Brown, M. Harada, J. W. Holt, M. Rho and C. Sasaki, arXiv:0804.3196 [hep-ph].

\bibitem{br91} G.E. Brown and M. Rho, Phys. Rev. Lett. {\bf 66} (1991) 2720.

\bibitem{HY:PR}  M. Harada and K. Yamawaki, Phys. Rept. {\bf 381} (2003) 1.

\bibitem{sakai} T.~Sakai and S.~Sugimoto,
  Prog.\ Theor.\ Phys.\  {\bf 114} (2005) 1083;
  Prog.\ Theor.\ Phys.\  {\bf 113} (2005) 843


\bibitem{mr:kyoto} M. Rho, arXiv:0805.3342 [hep-ph];  D.~K.~Hong, M.~Rho, H.~U.~Yee and P.~Yi,
 Phys.\ Rev.\  D {\bf 76} (2007) 061901; Phys.\ Rev.\  D {\bf 77} (2008) 014030.


\bibitem{mau} R. Vinh-Mau, J. M. Richard, B. Loiseau, M. Lacombe and
 W. N. Cottingham,
Phys. Lett. B {\bf 44} (1973) 1.

\bibitem{weise} T. Ericson and W. Weise, {\it Pions and Nuclei} (Clarendon
  Press, Oxford, 1988).

\bibitem{clayton} D. D. Clayton, {\it Principles of Stellar Evolution and
    Nucleosynthesis}
  (McGraw-Hill, New York, 1968).


\bibitem{BLR-STAR} G.~E.~Brown, C.~H.~Lee and M.~Rho,
  Phys.\ Rev.\  C {\bf 74} (2006) 024906.
\bibitem{phenix} S.~Afanasiev {\it et al.}  [PHENIX Coll.],
  arXiv:0706.3034 [nucl-ex].

\bibitem{footnotepp} We are counting the number of pions per
$\rho^0$ since we will normalize the lepton yield to the $\rho$
peak.
\bibitem{footnotep} These on-shell $\rho$'s are unaffected by the medium
with the free-space width $\Gamma_\rho=150$ MeV. They are included
in the ``cocktail" in the PHENIX data~\cite{phenix}. The same
argument holds in NA60 as shown in \cite{bhhrs}: the $\rho$
reconstructed by the NA60 assuming $T_{freezeout}=110$ MeV and
$m_\perp$ across the $\rho$ resonance is just the free $\rho$ with
no broadened width.


\bibitem{shuryak-brown} E.~V.~Shuryak and G.~E.~Brown,
  Nucl.\ Phys.\  A {\bf 717} (2003) 322.


\bibitem{PLB} H.-J. Park, C.-H. Lee and G.E. Brown, Nucl. Phys. {\bf A763} (2005) 197.

\bibitem{BLR-kaon} G.E. Brown, C.-H. Lee and M. Rho,
Phys.\ Rept.\  {\bf 462} (2008) 1.

\bibitem{halfskyrmion} M. Rho, arXiv:0711.3895 [nucl-th].

\bibitem{quarkyonic} L.~McLerran and R.~D.~Pisarski,
  Nucl.\ Phys.\  A {\bf 796} (2007) 83.

\bibitem{yamamoto}  T.~Hatsuda, M.~Tachibana and N.~Yamamoto,
  arXiv:0802.4143 [hep-ph].

\end{thebibliography}
\end{document}